\documentclass[prl,twocolumn,showpacs,preprintnumbers]{revtex4-1}
\usepackage{amsmath,bm,graphicx,hyperref}

\renewcommand\d{\partial}
\newcommand\grad{\bm{\nabla}}
\newcommand\+{\dagger}
\newcommand\<{\langle}
\renewcommand\>{\rangle}
\newcommand\p{{\bm{p}}}
\newcommand\q{{\bm{q}}}
\newcommand\ep{\epsilon_\p}
\newcommand\Ep{E_\p}
\newcommand\kF{k_\mathrm{F}}
\begin{document}
\preprint{LA-UR-12-23551}

\title{New type of crossover physics in three-component Fermi gases}

\author{Yusuke~Nishida}
\affiliation{Theoretical Division, Los Alamos National Laboratory,
Los Alamos, New Mexico 87545, USA}

\date{July 2012}

\begin{abstract}
 A three-component Fermi gas near a broad Feshbach resonance does not
 have a universal ground state due to the Thomas collapse, while it does
 near a narrow Feshbach resonance.  We explore its universal phase
 diagram in the plane of the inverse scattering length $1/a\kF$ and the
 resonance range $R_*\kF$.  For a large $R_*\kF$, there exists a
 Lifshitz transition between superfluids with and without an unpaired
 Fermi surface as a function of $1/a\kF$.  With decreasing $R_*\kF$, the
 Fermi surface coexisting with the superfluid can change smoothly from
 that of atoms to trimers (``atom-trimer continuity''), corresponding to
 the quark-hadron continuity in a dense nuclear matter.  Eventually,
 there appears a finite window in $1/a\kF$ where the superfluid is
 completely depleted by the trimer Fermi gas, which gives rise to a pair
 of quantum critical points.  The boundaries of these three quantum
 phases are determined in regions where controlled analyses are possible
 and are also evaluated based on a mean-field plus trimer model.
\end{abstract}


\maketitle

\section{Introduction}
The physics of a dilute two-component Fermi gas with a short-range
interaction becomes universal in the vicinity of a broad Feshbach
resonance.  In the limit of the vanishing potential range $r_0$ compared
to the $s$-wave scattering length $a$ and the Fermi wavelength
$\kF^{-1}$, the physics is characterized solely by their dimensionless
ratio $1/a\kF$.  With increasing $1/a\kF$, a two-component Fermi gas
exhibits a crossover from a Bardeen-Cooper-Schrieffer- (BCS-)type
superfluid to a Bose-Einstein condensate (BEC) of diatomic
molecules~\cite{Eagles:1969,Leggett:1980,Nozieres:1985}.  The BCS-BEC
crossover and its variations with mass and density imbalances have been
subject to extensive
research~\cite{Bloch:2008,Giorgini:2008,Proceedings,Lecture_Notes}.

This research field can be further extended to multicomponent Fermi
gases.  In particular, a three-component Fermi gas has attracted
considerable interest partially because of its intriguing analogy to a
matter of quarks with three
colors~\cite{Rapp:2007,Wilczek:2007,O'Hara:2011}, and, more recently,
because of its experimental realization with $^6$Li
atoms~\cite{Ottenstein:2008,Wenz:2009,Lompe:2010a,Lompe:2010b,Huckans:2009,Williams:2009,Nakajima:2010}.
Superfluid pairing and the BCS-BEC ``crossover'' of a three-component
Fermi gas have been studied
theoretically~\cite{Bedaque:2009,Paananen:2006,He:2006,Chang:2006,Zhai:2007,Paananen:2007,Cherng:2007,Catelani:2008,Floerchinger:2009,Martikainen:2009,Ozawa:2010,Nummi:2011,Salasnich:2011,Kanasz-Nagy:2012},
typically in mean-field approximations.

However, a serious problem arises when a many-body ground state of a
three-component Fermi gas is considered beyond the mean-field
approximation: Three distinguishable fermions can form an infinitely
deep bound state in the zero-range limit $r_0\to0$, which is known as
the Thomas collapse~\cite{Thomas:1935}.  Therefore, a three-component
Fermi gas does not have a many-body ground state, or even if it exists,
it is set by a nonzero $r_0$ and hence not universal.  This is in sharp
contrast to a two-component Fermi gas where the Thomas collapse does not
take place due to the Pauli exclusion principle.

\begin{figure}[b]
 \includegraphics[width=0.92\columnwidth,clip]{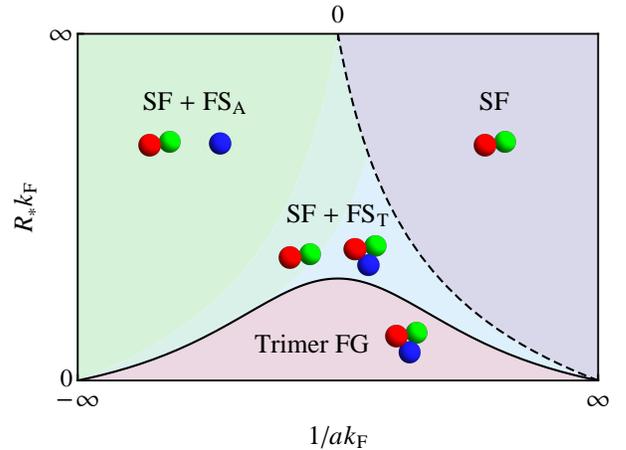}
 \caption{(color online).  Minimal phase structure of a three-component
 Fermi gas with equal masses, densities, and interactions at zero
 temperature.  There are three quantum phases consisting of superfluids
 (SF) with and without an unpaired Fermi surface and a Fermi gas of
 trimers (Trimer FG).  The Fermi surface coexisting with the superfluid
 can change smoothly from that of atoms (FS$_\mathrm{A}$) to trimers
 (FS$_\mathrm{T}$).  The phase boundaries are asymptotically given by
 Eqs.~(\ref{eq:superfluid_bcs}) and (\ref{eq:superfluid_bec}) for the
 superfluid transition (solid curve) and by
 Eqs.~(\ref{eq:lifshitz_narrow}) and (\ref{eq:lifshitz_bec}) for the
 Lifshitz transition (dashed curve).  See Fig.~\ref{fig:mean_field} for
 an evaluation based on a mean-field plus trimer model.
 \label{fig:phase_diagram}}
\end{figure}

The lack of a universal ground state of a three-component Fermi gas can
be cured in the vicinity of a narrow Feshbach resonance where both $a$
and the resonance range $R_*$ are much larger than
$r_0$~\cite{Petrov:2004}.  Because $R_*$ sets a ground state of three
fermions even in the zero-range limit $r_0\to0$, a dilute
three-component Fermi gas now becomes universal in the sense that its
physics is completely characterized by the two dimensionless ratios,
$1/a\kF$ and $R_*\kF$.  Recently, a two-component Fermi gas near a
narrow Feshbach resonance was studied experimentally using $^6$Li
atoms~\cite{Hazlett:2012} or $^6$Li-$^{40}$K
mixture~\cite{Kohstall:2012} as well as
theoretically~\cite{Ho:2012,Massignan:2012,Qi:2012,Trefzger:2012}.  Its
extension to three components of fermions is quite feasible.
Multicomponent Fermi gases can also be realized with alkaline-earth-like
atoms such as $^{173}$Yb~\cite{Taie:2010}.

The purpose of this work is to explore the universal phase diagram of a
three-component Fermi gas in the vicinity of a narrow Feshbach
resonance.  For simplicity, we shall consider the case of equal masses,
densities, and interactions~\cite{Zhang:2009,O'Hara:2011}, and our
findings are summarized in Fig.~\ref{fig:phase_diagram}.  There are
three regions in the plane of $1/a\kF$ and $R_*\kF$ where quantum phases
are easily located: a BCS superfluid with an unpaired Fermi surface in
the weak coupling limit $1/a\kF\ll-1$, a BEC superfluid with no unpaired
Fermi surface in the strong coupling limit $1/a\kF\gg1$, and a Fermi gas
of deeply bound three fermions (trimers) in the broad resonance limit
$R_*\kF\ll1$.  In what follows, their phase boundaries will be
determined in two regions where controlled analyses are possible: One is
the narrow resonance limit $R_*\kF\to\infty$ with fixed $1/a\kF$ (upper
side of Fig.~\ref{fig:phase_diagram}) and the other is the dilute limit
$\kF\to0$ with fixed $R_*/a$ (lower left and right corners).  A similar
approach was used in Ref.~\cite{Nishida:2010} to investigate the phase
diagram of a bilayer Fermi gas.  We also evaluate the phase boundaries
based on a mean-field plus trimer model (see Fig.~\ref{fig:mean_field})
and conclude that the minimal phase structure is already rich, as shown
in Fig.~\ref{fig:phase_diagram}.

\section{Narrow resonance limit}
A three-component Fermi gas near a narrow Feshbach resonance with
$\mathrm{SU}(3)\times\mathrm{U}(1)$ invariance is described by a
Lagrangian density (hereafter, $\hbar=1$):
\begin{align}
 \mathcal{L} &= \psi_i^\+\left(i\d_t+\frac{\grad^2}{2m}+\mu\right)\psi_i
 + \phi_i^\+\left(i\d_t+\frac{\grad^2}{4m}-\nu+2\mu\right)\phi_i \notag\\
 &\quad + g\,\frac{\epsilon_{ijk}}2
 \left(\phi_i^\+\psi_j\psi_k+\psi_k^\+\psi_j^\+\phi_i\right).
\end{align}
Here $\psi_i$ and $\phi_i$ represent fermionic atoms and bosonic
molecules, respectively, $\epsilon_{ijk}$ is the antisymmetric tensor,
and sums over repeated indices $i,j,k=1,2,3$ are assumed.  The bare
couplings $\nu$ and $g$ are related to the $s$-wave scattering length
and the resonance range by
\begin{align}
 \frac{\nu}{g^2} =  -\frac{m}{4\pi a} + \frac{m\Lambda}{2\pi^2}
  \qquad\text{and}\qquad g^2 = \frac{4\pi}{m^2R_*},
\end{align}
where the momentum cutoff $\Lambda$ should be sent to infinity.  A
controlled analysis is possible in the narrow resonance limit
$R_*\to\infty$ where the Feshbach coupling $g$ is vanishingly small and
thus the zero-temperature mean-field theory becomes
exact~\cite{Gurarie:2007}.

In order for all three components of fermions to have the same number
density $n_i\equiv\kF^3/(6\pi^2)$, three order parameters $\<\phi_i\>$
have to be equal up to arbitrary phases.  By introducing a gap parameter
$\Delta\equiv\sqrt3_{}g_{}|\<\phi_i\>|$, the grand potential density is
given by
\begin{align}\label{eq:mean_field}
 \Omega_\mathrm{MF}
 &= -\left(\frac{m}{4\pi a}+\frac{m^2\mu R_*}{2\pi}\right)\Delta^2
 - \frac{(2m\mu)^{5/2}}{30\pi^2m}\theta(\mu) \notag\\
 &\quad - \int\!\frac{d\p}{(2\pi)^3}\!
 \left[\Ep-(\ep-\mu)-\frac{\Delta^2}{2\ep}\right],
\end{align}
where $\ep\equiv\p^2/(2m)$, $\Ep\equiv\sqrt{(\ep-\mu)^2+\Delta^2}$, and
$\theta(\,\cdot\,)$ is the step function.  The second term is a
contribution of unpaired fermions and the last one is that of paired
fermions, both of which are superpositions of the original fermion
components $|i\>$.  For example, the unpaired fermion represented by a
blue particle in Fig.~\ref{fig:phase_diagram} is an equal superposition,
$|\mathrm{unpaired}\>=|1\>+|2\>+|3\>$.  This uniform state is
energetically favored over the phase separation discussed in
Refs.~\cite{Ozawa:2010,Titvinidze:2011}.  While the superfluid always
exists $\Delta\neq0$, the coexisting unpaired Fermi surface of atoms
appears only for $\mu>0$ and disappears for $\mu<0$.  Therefore, a
Lifshitz transition takes place at $\mu=0$~\cite{Bedaque:2009}.

By simultaneously solving the gap equation
$\d\Omega_\mathrm{MF}/\d\Delta=0$ and the number density equation
$3n_i=-\d\Omega_\mathrm{MF}/\d\mu$, a location of the Lifshitz
transition is found to be
\begin{align}\label{eq:lifshitz_narrow}
 R_*\kF = \frac{2^{10}\pi}{\Gamma(1/4)^8}(a\kF)^4
 - \frac{4}{3\pi}a\kF\,\bigg|_{a\kF>0}.
\end{align}
This equation determines the asymptotic behavior of the dashed curve in
Fig.~\ref{fig:phase_diagram} toward the narrow resonance limit
$R_*\kF\to\infty$.  We remind the reader that the zero-temperature
mean-field theory (\ref{eq:mean_field}) is correct only up to the
leading order of the systematic large $R_*\kF$
expansion~\cite{unpublished}.  Accordingly, the coefficient of the last
term in Eq.~(\ref{eq:lifshitz_narrow}) should be modified by
beyond-mean-field corrections.  The same Lifshitz transition has been
observed at $1/a\kF=0.633195$ in the broad resonance limit
$R_*\kF\to0$~\cite{He:2006,Ozawa:2010,Salasnich:2011}, but the
mean-field analysis breaks down here.

\section{Dilute limit}
\begin{figure}[t]
 \includegraphics[width=0.92\columnwidth,clip]{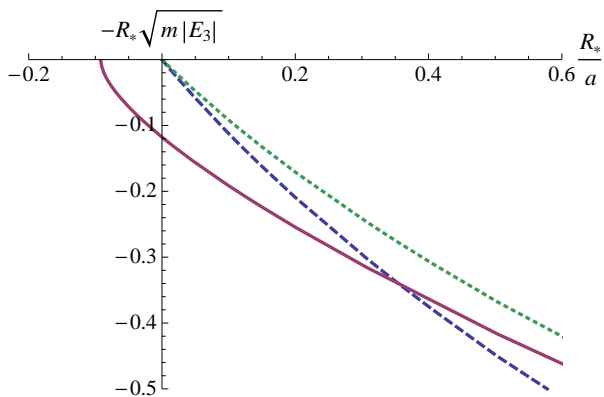}
 \caption{(color online).  Binding energy of a ground state trimer
 normalized as $-R_*\sqrt{m|E_3|}$ (solid curve).  The dotted and dashed
 curves are $-R_*\sqrt{m|E_2|}$ and $-R_*\sqrt{3m|E_2|/2}$,
 respectively, where $E_2$ is the dimer binding energy in
 Eq.~(\ref{eq:dimer}).  \label{fig:binding_energy}}
\end{figure}

A different controlled analysis is possible in the dilute limit
$\kF\to0$ where the problem reduces to a few-body problem.  In vacuum,
two distinguishable fermions form a bound state (dimer) with binding
energy $E_2<0$:
\begin{align}\label{eq:dimer}
 \sqrt{m|E_2|} = \frac{-1+\sqrt{1+4R_*/a}}{2R_*}\,\bigg|_{a>0}.
\end{align}
On the other hand, a binding energy of three distinguishable fermions is
determined by three coupled integral equations~\cite{Naidon:2011}:
\begin{align}\label{eq:integral_eq}
 & \left[\sqrt{\frac34\p^2+m|E_3|}-\frac1a
 + R_*\left(\frac34\p^2+m|E_3|\right)\right]\chi_i(\p) \notag\\
 &= \sum_{j\neq i}\int\!\frac{d\q}{(2\pi)^3}
 \frac{4\pi\chi_j(\q)}{\p^2+\q^2+\p\cdot\q+m|E_3|}.
\end{align}
Two types of solutions are potentially allowed: One is
$\chi_1(\p)=-\chi_2(\p)$ with $\chi_3(\p)=0$ where
Eq.~(\ref{eq:integral_eq}) becomes equivalent to a problem of
two-component fermions and hence no bound state solution; the other is
$\chi(\p)\equiv\chi_i(\p)$ for all $i=1,2,3$ where
Eq.~(\ref{eq:integral_eq}) becomes equivalent to a problem of three
identical bosons~\cite{Petrov:2004,Gogolin:2008}.  A ground state trimer
is found in the $s$-wave channel, $\chi(\p)=\chi(|\p|)$, whose binding
energy $E_3<0$ is plotted in Fig.~\ref{fig:binding_energy}.
While there is also an infinite tower of excited states due to the
Efimov effect~\cite{Efimov:1970}, they are irrelevant to the present
purpose to determine the many-body ground state.  We also note that
bound states formed with more than three fermions are unlikely due to
the Pauli exclusion principle.

With increasing $R_*/a$ on the side of $a<0$, the trimer appears from
the three-atom threshold at
\begin{align}\label{eq:superfluid_bcs}
 E_3 = 0 \quad\Leftrightarrow\quad \frac{R_*}a = -0.0917249.
\end{align}
On the left (right) side of this point, the dilute limit of a
three-component Fermi gas reduces to a system of atoms (trimers).
Because different components of fermions with equal densities can form
Cooper pairs by an infinitesimal attraction, this is actually a quantum
critical point to separate the atomic superfluid with an unpaired Fermi
surface from a Fermi gas of trimers.  Therefore,
Eq.~(\ref{eq:superfluid_bcs}) determines the asymptotic behavior of the
solid curve in Fig.~\ref{fig:phase_diagram} toward the dilute limit
$|a\kF|,R_*\kF\to0$.

We now turn to the side of $a>0$ but away from the point for
$E_3=E_2\ \Leftrightarrow\ R_*/a=2.18151$ where the trimer disappears
into the atom-dimer threshold.  In this region, the dilute limit of a
three-component Fermi gas reduces to noninteracting dimers and trimers
because their sizes become negligible compared to a mean interparticle
distance.  Since all dimers, if they exist, condense at zero
temperature, the chemical potential is fixed by their binding energy,
$2\mu=E_2$.  Then the total number density is a sum of contributions of
condensed dimers and trimers:
\begin{align}
 \frac{\kF^3}{2\pi^2} = n_\mathrm{dimer}
 + \frac{[6m(3\mu-E_3)]^{3/2}}{2\pi^2}\theta(3\mu-E_3).
\end{align}
With increasing $R_*/a$, dimers appear at $n_\mathrm{dimer}=0$, which
gives
\begin{align}\label{eq:superfluid_bec}
 E_3 + \frac{\kF^2}{6m} = \frac32E_2.
\end{align}
This is a quantum critical point to separate a Fermi gas of trimers on
its left side from the trimer Fermi gas coexisting with the dimer
superfluid on its right side.  With increasing $R_*/a$ further, the
Fermi surface of trimers disappears at
$n_\mathrm{dimer}=\kF^3/(2\pi^2)$, which gives
\begin{align}\label{eq:lifshitz_bec}
 E_3 = \frac32E_2 \quad\Leftrightarrow\quad \frac{R_*}a = 0.359011.
\end{align}
This is a Lifshitz transition, and the system beyond this point is the
dimer superfluid with no unpaired Fermi surface.  Therefore,
Eqs.~(\ref{eq:superfluid_bec}) and (\ref{eq:lifshitz_bec}) determine the
asymptotic behaviors of the solid and dashed curves in
Fig.~\ref{fig:phase_diagram}, respectively, toward the dilute limit
$a\kF,R_*\kF\to0$.

\section{Phase diagram}
Finally, we combine the above results to establish the universal phase
diagram of a three-component Fermi gas.  Because we developed controlled
understanding on quantum phases in all available limits of $1/a\kF$ and
$R_*\kF$, one emergent phase boundary has to end up as another phase
boundary of the same type.  Accordingly, the quantum critical lines for
the superfluid transition found in Eqs.~(\ref{eq:superfluid_bcs}) and
(\ref{eq:superfluid_bec}) are connected, and those for the Lifshitz
transition found in Eqs.~(\ref{eq:lifshitz_narrow}) and
(\ref{eq:lifshitz_bec}) are also connected.  This leads to the minimal
but rich phase structure shown in Fig.~\ref{fig:phase_diagram}.

An interesting observation is that the unpaired Fermi surface coexisting
with the superfluid is that of atoms in the weak coupling $1/a\kF\ll-1$
or dense region $R_*\kF\gg1$, while it is that of trimers in the strong
coupling and dilute region, $1/a\kF\gg1$ and $R_*\kF\ll1$.  Because
there is no sharp distinction between them in terms of symmetries or
topologies, the Fermi surfaces of atoms and trimers in the superfluid
phase can be smoothly connected.  If we associate three components of
fermions with three colors of
quarks~\cite{Rapp:2007,Wilczek:2007,O'Hara:2011}, this observation
corresponds to a smooth crossover from deconfined quarks to confined
baryons with decreasing density, which is known as the quark-hadron
continuity~\cite{Schafer:1999,Hatsuda:2006}.  Accordingly, the analogous
new feature in a three-component Fermi gas shall be termed an
``atom-trimer continuity.''

Regarding low-lying excitations in the superfluid phase, the breaking of
the $\mathrm{SU}(3)\times\mathrm{U}(1)$ symmetry down to
$\mathrm{SU}(2)\times\mathrm{U}(1)$ generates three Nambu-Goldstone
bosons~\cite{He:2006}.  Furthermore, in the presence of an unpaired
Fermi surface, there exists a gapless fermionic excitation which has the
character of an atom on one side and that of a trimer on the other side.
Because only the parity of a particle number is conserved in the
superfluid phase, an unpaired atom and a trimer cannot be distinguished
by their particle numbers.  Accordingly, the gapless fermionic
excitation can also exhibit a crossover with its effective mass changing
smoothly from $m$ to $3m$, which signals the atom-trimer
continuity~\cite{Powell:2005}.

\begin{figure}[t]
 \includegraphics[width=0.92\columnwidth,clip]{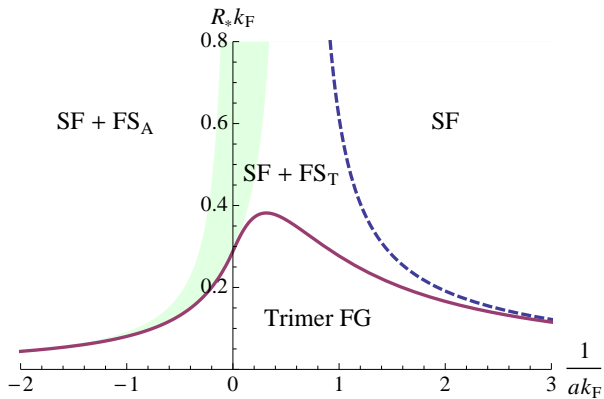}
 \caption{(color online).  Quantum critical lines for the superfluid
 transition (solid curve) and the Lifshitz transition (dashed curve)
 based on a mean-field plus trimer model (\ref{eq:model}).  The shaded
 region indicates where the smooth crossover from the Fermi surface of
 atoms (left side) to trimers (right side) takes place.
 \label{fig:mean_field}}
\end{figure}

In addition to the controlled analyses, we also develop a model analysis
to quantify the phase diagram.  A guiding principle to construct the
model is that it must incorporate the correct asymptotic behaviors
discussed above.  The simplest possibility is just to add a contribution
of noninteracting trimers to the mean-field grand potential density
(\ref{eq:mean_field}):
\begin{align}\label{eq:model}
 \Omega_\mathrm{MF+T} = \Omega_\mathrm{MF}
 - \frac{[6m(3\mu-E_3)]^{5/2}}{90\pi^2m}\theta(3\mu-E_3).
\end{align}
By solving the gap equation $\d\Omega_\mathrm{MF+T}/\d\Delta=0$ and the
number density equation $3n_i=-\d\Omega_\mathrm{MF+T}/\d\mu$ at
$\Delta\to0$, we find that the superfluid transition takes place at
\begin{align}
 E_3 + \frac{\kF^2}{6m} = \frac32E_2\,\theta(a).
\end{align}
The obtained quantum critical line is plotted in
Fig.~\ref{fig:mean_field} by the solid curve, which continuously
interpolates the correct asymptotic behaviors
[Eqs.~(\ref{eq:superfluid_bcs}) and (\ref{eq:superfluid_bec})] in the
weak and strong coupling limits $a\kF\to\pm0$.  In particular, we find
$R_*\kF=0.288325$ at $1/a\kF=0$ and its maximum $R_*\kF=0.381739$
reached at $1/a\kF=0.314545$.

On the other hand, in the range of Fig.~\ref{fig:mean_field}, the
Lifshitz transition takes place at $E_3=3\mu$ where the Fermi surface of
trimers disappears with increasing $1/a\kF$.  The obtained quantum
critical line is plotted by the dashed curve, which again yields the
correct asymptotic behavior [Eq.~(\ref{eq:lifshitz_bec})] in the strong
coupling limit $a\kF\to+0$.  Similarly, the Fermi surface of trimers
appears at $E_3=0$ and that of atoms disappears at $\mu=0$, which
correspond to the left and right edges of the shaded region in
Fig.~\ref{fig:mean_field}, respectively.  While they define sharp
boundaries in our simple treatment of trimers as noninteracting
particles, such sharp boundaries need not actually exist.  Indeed, a
more elaborate model with interaction terms
$\sim t^\+\psi_i\phi_i+\mathrm{H.c.}$ added to the mean-field plus
trimer model can incorporate the atom-trimer continuity by hybridizing
an unpaired atom and a trimer (represented by $t$) by a condensed
dimer~\cite{Powell:2005}.  The shaded region is thus meant to indicate
where the smooth crossover from the Fermi surface of atoms to trimers
takes place with increasing $1/a\kF$.

\begin{figure}[t]
 \includegraphics[width=0.92\columnwidth,clip]{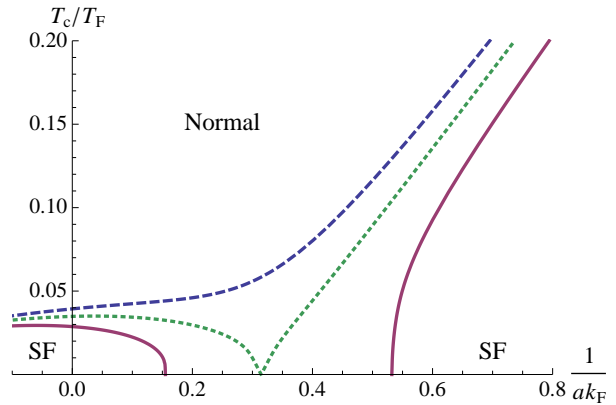}
 \caption{(color online).  Superfluid critical temperatures in units of
 $T_\mathrm{F}\equiv\kF^2/(2mk_\mathrm{B})$ for $R_*\kF=0.40$ (dashed
 curve), $0.381739$ (dotted curve), and $0.36$ (solid curve).  These
 values of $R_*\kF$ are above, at, and below the critical resonance
 range for opening the complete depletion of the superfluid by the
 trimer Fermi gas, which gives rise to a pair of quantum critical
 points.  \label{fig:Tc}}
\end{figure}

We also extend the mean-field plus trimer model (\ref{eq:model}) to
nonzero temperature with a caveat that the mean-field approximation
overestimates the critical temperature in the strong coupling region
$1/a\kF\gtrsim0$~\cite{Melo:1993}.  Figure~\ref{fig:Tc} shows the
obtained superfluid critical temperature as a function of $1/a\kF$ for
three different values of $R_*\kF$.  With decreasing $R_*\kF$ from
above, the superfluid is gradually suppressed by the emergence of
trimers.  Eventually, below the critical value $R_*\kF=0.381739$, there
appears a finite window in $1/a\kF$ where the superfluid is completely
depleted by the trimer Fermi gas, which gives rise to a pair of quantum
critical points.  These are the same type of quantum critical point as
that studied in the context of Bose-Fermi
mixtures~\cite{Powell:2005,Fratini:2010}.  Also, the same role of the
trimer in the dilute limit was discussed in
Ref.~\cite{Floerchinger:2009} based on approximate three-body
calculations.

\section{Summary and discussion}
A narrow Feshbach resonance sets the stage to investigate universal
aspects of a three-component Fermi gas.  We explored its universal phase
diagram with equal masses, densities, and interactions and found the
minimal but rich phase structure shown in Fig.~\ref{fig:phase_diagram}.
Our main discovery is a new type of crossover physics: In addition to
the ordinary BCS-BEC crossover from loosely bound Cooper pairs to
tightly bound dimers, unpaired fermions coexisting with the superfluid
can change smoothly from atoms to trimers, corresponding to the
quark-hadron continuity in a dense nuclear
matter~\cite{Schafer:1999,Hatsuda:2006}.  This new feature, termed an
atom-trimer continuity, provides a novel analogy between ultracold atoms
and quantum chromodynamics and should be investigated further.

At zero temperature, even identical fermions can form Cooper pairs in
some partial wave channel~\cite{Kohn:1965}.  An interaction between
unpaired atoms in the weak coupling region $1/a\kF\ll-1$ is induced by a
density fluctuation of the other components of fermions, while an
interaction between unpaired trimers in the strong coupling region
$1/a\kF\gg1$ is induced by a density fluctuation of the superfluid
dimers which dominates over a direct trimer-trimer interaction.  Because
both induced interactions cause an instability in the $p$-wave
channel~\cite{Bulgac:2006}, the previously unpaired Fermi surface
exhibits the $p$-wave superfluidity regardless of whether it was of
atoms or trimers.  Therefore, the atom-trimer continuity remains intact.
Here the $p$-wave pairing of atoms or trimers breaks the
$\mathrm{SU}(2)\times\mathrm{U}(1)$ symmetry of the $s$-wave superfluid
phase down to $\mathrm{SU}(2)\times\mathrm{Z}_2$ besides broken
rotational symmetries.  On the other hand, because the $s$-wave
superfluid phase with no unpaired Fermi surface retains the
$\mathrm{SU}(2)\times\mathrm{U}(1)$ symmetry, the previous Lifshitz
transition (dashed curve in Fig.~\ref{fig:phase_diagram}) becomes the
$p$-wave superfluid transition.  The trimer Fermi gas phase is still
distinct from the other two phases because its symmetry breaking pattern
is different,
$\mathrm{SU}(3)\times\mathrm{U}(1)\to\mathrm{SU}(3)\times\mathrm{Z}_6$
by any pairing of trimers.

Thus far we focused on the maximally symmetric case, while our approach
can be easily generalized to unequal masses, densities, and
interactions.  Furthermore, the ideas developed here can be used to
investigate universal aspects of other systems, such as a two-component
Fermi gas with a large mass ratio $>13.6$~\cite{Petrov:2003}, which also
lacks a universal ground state in the vicinity of a broad Feshbach
resonance.

\acknowledgments
The author thanks J.~Carlson, S.~Reddy, and N.~Yamamoto for valuable
discussions and a LANL Oppenheimer Fellowship for the support of this
work.

\end{document}